# Spectroscopy based on target luminescence caused by interaction with ultrashort UV laser pulses

A A Ionin[1], D V Mokrousova[1,2], L V Seleznev[1], D V Sinitsyn[1], E S Sunchugasheva*[1,2] and N A Fokina[2]

[1] Lebedev Physical Institute, Russian Academy of Sciences, 53 Leninskiy prospect, 119991 Moscow, Russia
[2] Moscow Institute of Physics and Technology, 9 Institutskiy Pereulok, 141701 Dolgoprudny, Moscow Region, Russia
*Corresponding author: ses@lebedev.ru

We proposed remote spectroscopy approach consisted in luminescence light utilization. During interaction with different targets ultrashort UV laser pulse generates broadband spectrum light, which can be applied for remote spectroscopy purposes. We selected appropriate target materials to cover required spectral range from 300 to 600 nm and provided an example of spectrum reconstruction of known material. Obtained spectra are in a good correlation with calculated ones.

Ultrashort laser pulses are widely applied in remote spectroscopy (for example, see review [1]). Two approaches are implemented in remote spectroscopy: supercontinuum-based and laser induced spectroscopy. The first one utilizes supercontinuum light that is generated during filamentation of high-power laser pulses. Main advantages of this procedure are broad spectrum and short duration of supercontinuum pulses. In the other procedure plasma is generated due to target surface ablation by intense laser pulses, and one detects plasma luminescence spectrum. Both approaches need high intensity of the laser pulse at a remote target or determined space at some distance from the laser, which can be obtained with high-power laser pulse filamentation taking place under its propagation in the air These approaches have been mainly realized for IR ultrashort pulses. Ultrashort UV pulse with low energy but with high energy of its quantum and high peak intensity can be transmitted through a long distance via filamentation and allow one to obtain bright luminescence without surface ablation. In this paper we study luminescence of a target caused by its interaction with ultrashort UV laser pulses and an opportunity of using this luminescence as a pulse source for spectroscopy purposes.

An optical scheme of the experiment is depicted in fig.1. UV laser pulse at central wavelength of 248 nm, pulse duration of 100 fs and energy of (120±15) μJ illuminated the target. Luminescent light from this target was detected by a spectrometer via focusing lens with diameter of 4 cm and focal distance of 4 cm. The distance between the target and focusing lens was up to 1 m.

Luminescent spectra of three different targets are shown in fig.2. In the experiment we used stained glasses ZhS19 and BS11 and white office paper as targets. Absorbing media was modelled by inserting additional stained glasses ZS7 and ZS8 between the luminescence source and spectrometer. Luminescence spectrums of ZhS19 with additional glass ZS7 and without it are represented in fig.3. Obtained spectra allowed us to reproduce spectral characteristics of absorbing media. Fig.4. shows spectral transmittance of stained glass ZS7 and ZS8 measured in the experiments and compared with calculated data via GOST (EASC) tables for these glasses. From this figure one can see that experimental and numerical data are in a good quantitative correlation. Inaccuracy of transmittance measuring in the vicinity of the luminescence spectral edge increased due to low initial (before absorption) intensity of luminescence.

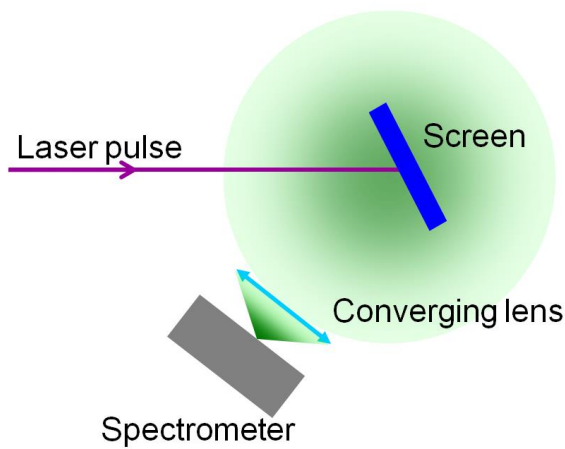

Figure 1 Optical scheme of the experiment

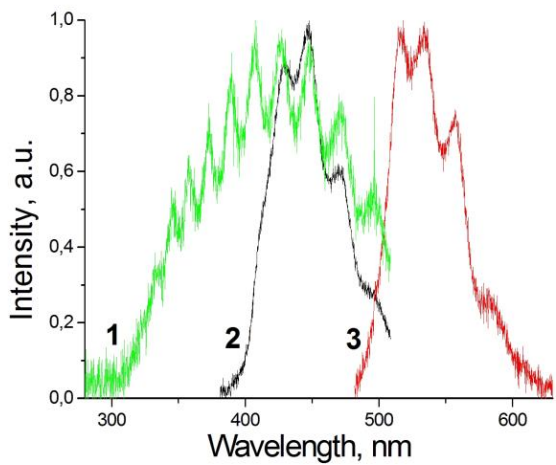

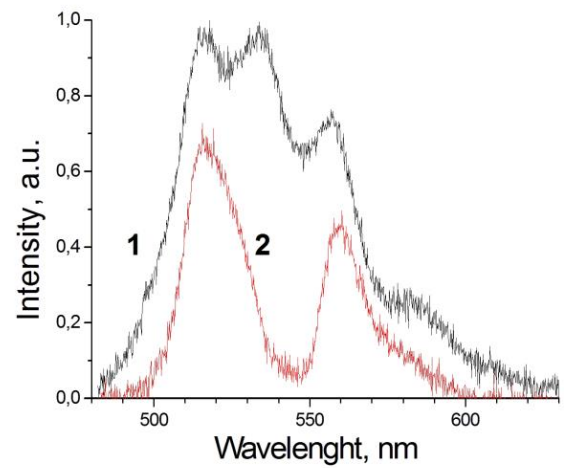

Figure 2 Target luminescence spectra: 1 – BS11, 2 – paper, 3 – ZhS19

Figure 3 Luminescence spectrum of the target ZhS19 without (1) and with (2) absorbing glass stained glass ZS7.

Nowadays there are laser systems that generate high-power ultrashort UV laser pulses (for example, see [2]). The system in Ref. [2] allows one to obtain a single ultrashort UV laser pulse or a train of such pulses of sub-TW power that is several orders of magnitude higher than the critical power of self-focusing for UV pulses. Energy density distribution over transverse cross-section of the beam at the distance of 50 m from the laser amplifier is shown in fig.5. Black dots correspond to hot-spots, i.e. to positions of high-intense filaments. So, it is clear that the laser pulse propagates in multi-filament regime. Based on the data from [2], we demonstrated an opportunity for the laser beam to propagate with relatively large cross-section of several centimeters with high-intensity filaments. Therefore, if we put luminescent screen into the beam, we can obtain a bright source of broad-spectrum light applicable for spectroscopy purposes. Actually, the photograph in Fig. 5 was registered by a CCD-camera directed to a luminescent glass plate. Despite the high-intensity filaments in the beam, the plate surface degradation did not take place in that experiment because of low energy of the UV filaments.

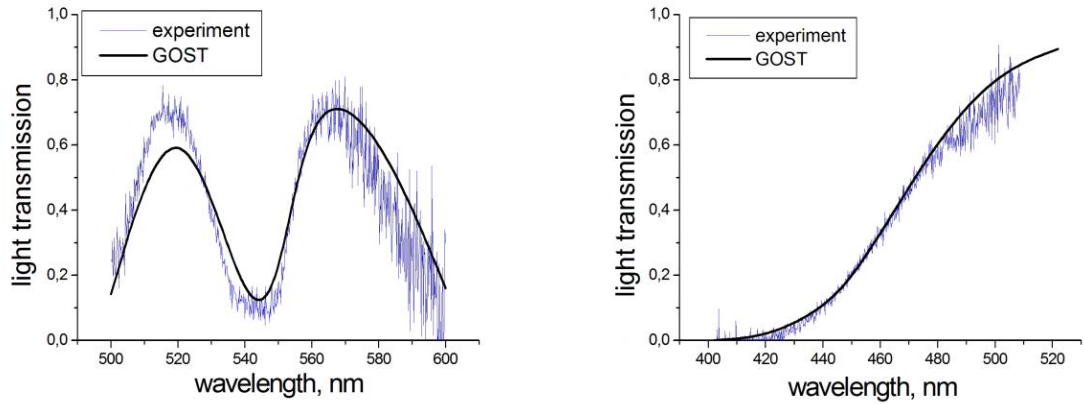

Figure 4 Transmittance of stained glass ZS7 (left) and ZS8 (right): experimentally obtained and calculated from GOST (EASC) standard.

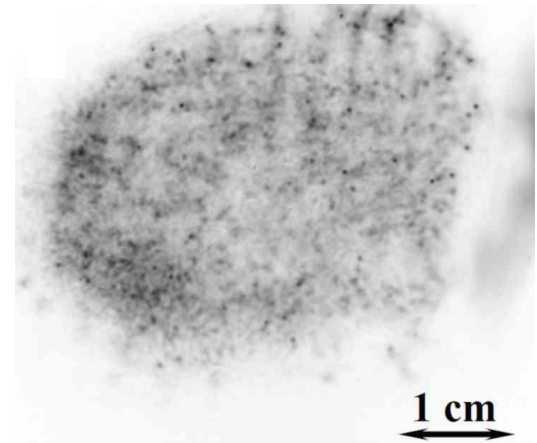

Figure 5 Laser beam cross-section at 50 m distance from KrF laser amplifier [2].

To sum up, in this paper we suggested and realized the procedure of absorption spectra measurement consisting in using target luminescence during interaction with ultrashort UV laser pulses as a source of probe light. We obtained high-intensity UV laser beam with relatively large diameter at significant distance from laser system in multifilament regime. Using of different target material allowed us to get broad spectrum luminescence (in this paper we used light with wavelengths from 300 to 600 nm). We demonstrated an opportunity of application of such a broad-spectrum luminescence source for spectroscopy purposes.

This research was supported by grants: RFBR 14-02-00489, 14-22-02021, LPI Educational-Scientific Complex, grant of the President of the Russian Federation NSh-3796.2014.2.

1. Kandidov V P, Slenov S A, Kosareva O G 2009 Quantum Electronics 39 205
2. Ionin A A, Kudryashov S I, Levchenko A O, Seleznev L V, Shutov A V, Sinitsyn D V, Smetanin I V, Ustinovsky N N, and Zvorykin V D 2012 AIP Conference Proceedings 1464 711